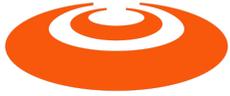
**Whitepaper**

# Exploring the intersection of Generative AI and Software Development

Filipe Calegario[1], Vanilson Burégio[2], Francisco Erivaldo[3], Daniel Moraes Costa Andrade[1], Kailane Felix[1], Nathalia Barbosa[1], Pedro Lucas da Silva Lucena[1], César França[3]

[1]Centro de Informática - UFPE
Recife, Brazil

[2]Universidade Federal Rural de Pernambuco - UFRPE
Recife, Brazil

[3]CESAR Innovation Center
Recife, Brazil

fcac@cin.ufpe.br
vanilson.buregio@ufrpe.br
franssa@cesar.school

## Abstract

In the ever-evolving landscape of Artificial Intelligence (AI), the synergy between generative AI and Software Engineering emerges as a transformative frontier. This whitepaper delves into the unexplored realm, elucidating how generative AI techniques can revolutionize software development. Spanning from project management to support and updates, we meticulously map the demands of each development stage and unveil the potential of generative AI in addressing them. Techniques such as zero-shot prompting, self-consistency, and multimodal chain-of-thought are explored, showcasing their unique capabilities in enhancing generative AI models. The significance of vector embeddings, context, plugins, tools, and code assistants is underscored, emphasizing their role in capturing semantic information and amplifying generative AI capabilities. Looking ahead, this intersection promises to elevate productivity, improve code quality, and streamline the software development process. This whitepaper serves as a guide for stakeholders, urging discussions and experiments in the application of generative AI in Software Engineering, fostering innovation and collaboration for a qualitative leap in the efficiency and effectiveness of software development.

# Introduction

The age of Artificial Intelligence (AI) has opened up new avenues for innovation across all industries, and Software Engineering is no exception. While the application of AI in many areas has shown impressive results, the intersection between generative AI and software development is still a largely unexplored area. This whitepaper addresses this gap by exploring how generative AI techniques can be applied to improve productivity, accelerate developer learning, and improve overall software quality.

In the field of Software Engineering, we believe it is essential to map the demands in all stages of the software development lifecycle, from management to support and updates. By addressing each aspect - Management, Requirements, UX/UI, System Architecture, Database, Coding, Testing, Deployment, Operation and Maintenance, and Support and Updates - we hope to identify opportunities for effectively applying generative AI.

Generative AI has the potential to revolutionize the way we think about software development. Training and fine-tuning can teach AI to understand and replicate patterns in code, while prompt engineering can translate human needs into instructions for the AI. Semantic search allows AI to understand and respond to queries more naturally, while AI agents can take on repetitive or low-value tasks, freeing humans to focus on more complex and creative tasks. AI tools and plugins can be integrated into existing platforms, making AI more accessible to developers, and code assistants can provide real-time help to speed up the development process.

## Generative AI and Software Development

In this whitepaper, we have explored the intersection between Generative AI and Software Development, specifically focusing on the potential demands of software engineering in different stages of software development and the potential Generative AI techniques that can be applied.

We began by discussing the various stages of software engineering, including management, requirements, UX/UI design, system architecture, database, coding, testing, deployment, operation and maintenance, and support and updates. We highlighted the common aspects and artifacts associated with each stage, providing a comprehensive overview of the software development process.

Next, we delved into the potential demands of software engineering that can be addressed using Generative AI techniques. We discussed techniques such as automatically generating project backlogs, business documentation, high-level specifications, requirement documents, wireframes and prototypes, descriptive text for interfaces, generating high- and low-level diagrams, optimizing database design and queries, assisting in data migration and code optimization, generating code snippets and prototypes, automated test report generation, deployment automation, chatbots for user support, and generating responses to incidents and customer feedback.

## Generative AI Techniques for Software Development

We also explored the potential Generative AI techniques that can be applied in software development. These techniques include zero-shot prompting, few-shot prompting, chain-of-thought prompting, self-consistency, generate knowledge prompting, tree of thoughts, automatic reasoning and tool-use, and multimodal chain-of-thought. Each technique offers unique capabilities to enhance the performance and functionality of Generative AI models in software development tasks.

Furthermore, we discussed the importance of vector embeddings and context in Generative AI models. Vector embeddings provide a representation of words in a multidimensional space, capturing semantic information and enabling similarity calculations. Context plays a crucial role in understanding and generating coherent and relevant text, allowing Generative AI models to consider the broader context of a task or query.

We also highlighted the significance of plugins, tools, LLM agents, and code assistants in enhancing the capabilities of Generative AI models in software development. These components provide additional functionality, optimization, visualization, and integration with external APIs, enabling developers to leverage the full potential of Generative AI in their software development workflow.

## Embracing the Future: Generative AI's Role in Advancing Software Development

The intersection between Generative AI and Software Development offers immense potential for addressing the demands and challenges of software engineering. By applying Generative AI techniques and leveraging advanced tools and plugins, developers can enhance their productivity, improve code quality, and streamline the software development process. The future of Generative AI in software development is promising, and further research and innovation in this field will continue to drive advancements in the software engineering domain.

This whitepaper is intended for researchers, software engineers, IT managers and other stakeholders in the Information and Communication Technologies ecosystem, with a special focus on universities and software companies located in the Porto Digital in Recife, Pernambuco, Brazil. We hope this research stimulates more discussions and experiments around the use of generative AI in Software Engineering, paving the way for future advances in the area.

As we delve into the possibilities of this promising intersection, we hope to shed light on new ways of thinking and approaching the challenges of Software Engineering, thus promoting a qualitative leap in the efficiency and effectiveness of software development. We believe that, with the support of everyone involved in the Information and Communication Technologies ecosystem, we can open new frontiers and achieve results that seem, at the moment, futuristic, but are certainly tangible at the confluence between generative AI and software development.

# SOFTWARE ENGINEERING STAGES

Software development is a process that encompasses a wide range of activities, from the initial idea conception to the maintenance of the software after delivery. Software development methodologies are structured approaches that guide this process, defining the steps and activities that take a software project from start to finish. Several factors, such as the project size, available resources, and preferences of the development team and the client, among other aspects influence the choice of a development methodology.

Several methodologies are commonly used in software development, some of which include:

- **Waterfall Development:** This classic model encourages a linear, sequential approach to development, where each phase must be completed before proceeding to the next. Contrary to popular belief, the Waterfall model allows for revision and revisiting of previous phases when necessary.
- **Iterative and Incremental Development:** In this approach, the software is divided into smaller parts or increments, which are developed in successive iterations. This allows for greater flexibility and the possibility of receiving constant feedback during development.
- **Agile Development:** As a response to the perceived rigidity in the waterfall model, Agile Development emphasizes continuous collaboration between the development team and the customer, adaptability to change, and frequent delivery of working software. Several agile frameworks exist, including Scrum, Kanban, and Extreme Programming (XP).
- **Lean Development:** With origins in the Toyota Production System, Lean Development focuses on eliminating waste, maximizing customer value, and continuous software delivery.
- **Model-Driven Software Development (MDSD):** This model places emphasis on creating high-level abstract models that describe the structure and behavior of software, thus facilitating better abstraction and automation of the development process.
- **Spiral Development:** This model combines elements of waterfall and iterative methodologies, focusing on risk analysis and mitigation.
- Component-Oriented Software Development (CBD): This model promotes the reuse of existing software components in building new systems, increasing efficiency and modularity.
- **Test Driven Development (TDD):** In this paradigm, tests are written before code, and software is developed to pass these tests. This helps ensure code quality and catch defects early.

## Aspects of development

Although there are specific nuances for each development methodology, we can identify common aspects present in most software projects. It is worth mentioning that the names and organization of the artifacts produced may vary according to the methodology employed (e.g. Waterfall, Agile) and the project's specific requirements. Furthermore, in more agile approaches, these phases and their respective artifacts can be iterative, due to the cyclical and incremental nature of these methodologies.

In this whitepaper, phases and artifacts are grouped into the following aspects: Management, Requirements, UX/UI, System Architecture, Database, Coding, Testing, Deployment, Operation and Maintenance, and Support and Updates.

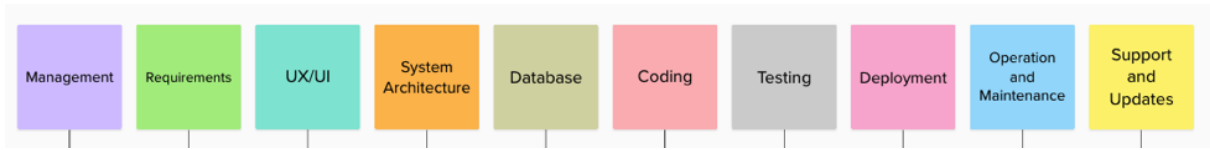

**Figure 1** - The group of aspects related to the software development process.

1. **Management**: Encompasses the planning, coordination, measurement, monitoring, and control of all activities that occur during the software development lifecycle. The main objective is to ensure the software is delivered on time, within budget, and according to the agreed specifications and requirements.
2. **Requirements**: This phase involves gathering, analyzing, documenting, and validating system and user requirements. It is vital to ensure that the final product meets the user's needs and expectations.
3. **UX/UI**: During this phase, user interface (UI) and user experience (UX) design elements are created. It includes planning how the end user will perceive and use the software, ensuring its accessibility and ease of use.
4. **System Architecture**: Involves defining the overall structure of the software system, including the selection of design patterns, quality attributes, and design decisions that include, among other things, defining how the different components of the software interact with each other.
5. **Database**: In this phase, data storage requirements and strategies for its management are defined. This involves designing the database schema, selecting the appropriate database management system, and implementing data security measures.
6. **Coding**: In this phase, software designs and architectures are translated into code, using appropriate programming languages and following best coding practices.
7. **Tests**: The software undergoes a series of tests to identify and correct errors, verify functionality, and ensure that it meets specified requirements.
8. **Deployment**: The software is installed in the end-user environment. May involve activities such as installation, configuration, performance optimization and others
9. **Operation and Maintenance**: After deployment, it is necessary to maintain the software to ensure its continued operation. This may include bug fixing, performance improvements, and updating the software to meet new requirements or changes to the runtime environment.
10. **Support and Updates**: This phase involves providing user support, fixing software issues, and providing updates and enhancements requested by users.

It is important to highlight that while these phases might have different names in various development methodologies, they usually correspond to standard steps in many software development processes. We have presented these steps in a linear order, but in real-world scenarios, some steps may overlap or be revisited, particularly in agile methodologies. The best way to integrate these steps is influenced by the specific needs of the project and the guidelines and practices followed by the organization and its development team. In the following sections, we will examine more closely the typical outputs or artifacts produced at each stage of the software development process.

## Software Artifacts: Beyond Source Code

While source code is a key component of any software project, it is not the sole artifact. Software artifacts encompass all materials produced throughout a project's duration, from the initial requirements documentation to project plans, and from user manuals to maintenance reports.
These artifacts are instrumental in the creation and upkeep of a software system. Their importance extends beyond the immediate project as many can be repurposed, reducing effort and time in subsequent projects.

Consider this: a detailed requirements document lays the groundwork for the project, steering the following design and development stages. User interface drafts ensure the software remains user friendly. Diagrams outlining system architecture give developers a clear picture of the software's structure. Automated tests consistently verify the software's functionality, while supplementary documents aid users in effectively utilizing the software.

To summarize, every software artifact holds significant value, contributing to the end goal of a top-tier software product. Table 1 outlines the frequent phases of the software development process discussed in this article, highlighting typical artifacts produced during each phase.

| Aspect | Description of Artifacts | Examples of Artifacts |
| --- | --- | --- |
| Management | Documents that guide the planning, coordination, monitoring and control of the project | Project Plan, Project Schedule, Progress Reports |
| Requirements | Documents that define what the software should do | Requirements Document, Use Case Diagrams, User Stories |
| UX/UI Design | UI and UX design elements | Wireframes, User Interface Prototypes, Interface Design Specifications |
| System Architecture | Documents that define the overall structure of the software system | Software Architecture Document, Component Diagrams, Sequence Diagrams |
| Database | Documents and schemas that define how data is stored and managed | Entity-Relationship Diagrams, Database Schemas, Database Creation Scripts |
| Codification | Source code and related documentation | Source Code, Code Documentation, Compilation Scripts |
| Tests | Documents, and scripts used to | Test Plans, Test Cases, Test |

| | test software functionality and robustness | Scripts, Bug Reports |
|---|---|---|
| Implantation | Scripts, and documentation used to install the software in the user's environment | Deployment Scripts, CI/CD Pipelines, Configuration Instructions, Container Definitions (e.g., DockerFIles) |
| Operation and maintenance | Documents and records used to monitor and maintain the software | Incident Reports, Change Records, Maintenance Documentation |
| Support and Updates | Documents and guides used to assist users and provide software updates | User Guides, FAQ, Update Versions Documentation |

**Table 1** - Artifacts associated with the phases of the software development process

# POTENTIAL SOFTWARE ENGINEERING DEMANDS FOR GENERATIVE AI

In this section, we explore various Software Engineering situations that can greatly benefit from the application of Generative AI techniques. Our team has worked hard to gather a wide range of potential use cases, using information from many online sources. These sources include recent tweets, blog posts, and academic papers on arXiv, all collected between March and July 2023.

The Generative AI field is always changing and growing. So, while our detailed review covers many new methods, tools, and applications, it only shows a moment in time of this fast-moving area. The strong connection we see between Generative AI and different areas of software development will continue to change and get stronger. Readers should see this list as a starting point that can be updated and made better as more information becomes available.

## Management

- **Automatically generate project backlogs**: Utilizing Generative AI, a project backlog can be systematically formulated by assessing project objectives, historical project data, or feedback from stakeholders. Serving as a prioritized enumeration of tasks, functionalities, or requirements, the backlog enhances the efficiency of the planning phase.
- **Business documentation**: By harnessing Generative AI, business documentation can be auto-generated. This ensures a comprehensive encapsulation of vital elements like business goals, stakeholder roles, and the scope of the project, aligning all stakeholders and clarifying project aims.
- **High-level specifications**: Generative AI can transform foundational ideas or overarching project visions into high-level specifications. This involves delineating the general system architecture, primary components, and their interconnections, offering a clear system overview to developers and stakeholders.
- **Requirement documents**: By interpreting stakeholder interviews, user feedback, or preliminary drafts, Generative AI can produce detailed requirement documents autonomously. These documents stand as a definitive guide for developers, detailing the software's objectives, required features, and inherent constraints.

## Requirements

- **Create requirement docs from a high-level input**: Utilizing Generative AI techniques, it's possible to transform high-level, often abstract, project inputs into detailed requirement documents. This process ensures that broad project visions or stakeholder objectives are distilled into actionable, clear, and structured software requirements, providing developers with a concrete foundation upon which to build.
- **Convert informal user feedback and comments into structured software requirements**: Generative AI can systematically interpret and process informal feedback, comments, or even anecdotal user experiences, converting them into structured software requirements. This capability allows developers to directly incorporate real-world user insights into the software design, enhancing its relevance and usability.

## UX/UI Design

- **Generate wireframes and prototypes based on software requirements**: Through advanced Generative AI tools, it's feasible to derive wireframes and interactive prototypes directly from software requirements. This automation not only expedites the design phase but also ensures that the design aligns closely with the stipulated requirements, bridging the gap between functional necessities and visual representation.
- **Generate descriptive text for interfaces and prototypes**: Generative AI can craft contextually relevant and user-friendly textual content for interfaces and prototypes. By analyzing the UI elements and their intended functionality, the AI can provide descriptive text, tooltips, and labels, enhancing the user's comprehension and interaction with the interface.

## System Architecture

- **Generate high- and low-level diagrams from requirements or textual descriptions**: Employing Generative AI tools, system architectures can be visualized by automatically transforming software requirements or textual descriptions into high-level (overview) and low-level (detailed) diagrams. This automation streamlines the transition from abstract requirements to concrete architectural blueprints.
- **Create system design documents by translating requirement documents**: Generative AI can facilitate the generation of detailed system design documents by interpreting and converting existing requirement documents. This ensures a seamless transition from the "what" (requirements) to the "how" (design and architecture) in software development.
- **Generate migration guidelines and representations for system animations and workflow diagrams**: By analyzing current system states and desired endpoints, Generative AI can produce migration guidelines, assisting teams in transitioning between software versions or platforms. Additionally, it can generate dynamic system animations and workflow diagrams, providing visually engaging representations of system processes and interactions.
- **Predict potential system bottlenecks and provide architecture optimization recommendations**: Leveraging advanced analytics and pattern recognition, Generative AI can preemptively identify potential system bottlenecks or inefficiencies. Based on these predictions, it can then proffer recommendations to optimize the system architecture, ensuring smoother and more efficient system operations.

## Database

- **Provide a data modeling assistant for efficient database design**: Using Generative AI, a virtual assistant can be developed to aid in the data modeling process. This assistant can offer suggestions, ensure normalization, and help in constructing efficient database designs that cater to the specific needs of an application.
- **Auto-generate and streamline database queries**: Generative AI can automatically craft precise database queries based on user intent or high-level descriptions. This not only speeds up the development process but also minimizes potential human errors, ensuring that data retrieval and manipulation are both efficient and accurate.
- **Process and transform complex data schemas, resolving inconsistencies**: Generative AI tools can analyze intricate data schemas, identify inconsistencies or redundancies, and subsequently resolve them. This capability is crucial for maintaining data integrity and coherence, especially in systems with large or evolving datasets.
- **Optimize and streamline database queries and procedures**: Through pattern recognition and benchmarking, Generative AI can review existing database queries and procedures, suggesting optimizations. This can lead to faster query responses, reduced server loads, and an overall more responsive database system.

## Codification

- **Assist in data migration and code optimization**: Generative AI can serve as an invaluable assistant when migrating data between systems or optimizing existing code. By analyzing the source and destination systems, it can ensure seamless transitions and identify areas in code that can be enhanced for performance and clarity.
- **Automatically produce code snippets, complete programs, and regular expression patterns based on requirements**: Through advanced analysis of provided requirements, Generative AI can auto-generate specific code snippets, entire software programs, or even intricate regular expression patterns. This accelerates development and ensures that the generated output adheres closely to the stipulated requirements.
- **Refactor and enhance existing code with suggestions for feature integration**: By examining current codebases, Generative AI can suggest refactoring strategies to improve code readability and performance. Additionally, it can recommend new features or integrations that can augment the functionality and value of the software.
- **Generate code prototypes based on high-level requirements**: Given a set of high-level requirements or abstract project objectives, Generative AI can quickly draft code prototypes. These prototypes can serve as foundational blocks for further development, providing a tangible starting point that aligns with the project's goals.
- **Translate code between programming languages, optimizing syntax and structure for the target language**: Generative AI can act as a code translator, converting code from one programming language to another. While doing so, it ensures that the translated code is not just syntactically correct but also optimized for performance and readability in the target language.
- **Convert code into interactive diagrams and visual representations**: By parsing through code structures and logic, Generative AI can transform raw code into interactive diagrams or visual aids. These representations can be invaluable for understanding complex algorithms, workflows, or data structures, aiding both developers and stakeholders.

## Tests

- **Automated test report generation and test creation suggestions**: Generative AI can be utilized to automatically produce detailed test reports post-execution, highlighting pass rates, failures, and anomalies. Furthermore, it can proactively suggest the creation of specific tests based on observed patterns and potential vulnerabilities in the code. • Generate test cases and suites, mock components, and restructure code for optimization: Generative AI can derive comprehensive test cases and suites from requirements, ensuring thorough coverage. In addition, it can fabricate mock components for isolated testing and suggest code restructuring to improve testability and performance.
- **Collaborative code analysis tool that identifies areas for improvement and offers feedback**: Integrating collaborative features with AI-driven code analysis, this tool can scan code repositories in real-time during team collaborations, pinpointing areas for improvement and providing instant feedback to developers.
- **Identify and rectify syntax errors, security vulnerabilities, and suggest mitigations**: Generative AI can act as an advanced linter and security auditor, detecting and fixing syntax errors, and identifying potential security risks in the code. For each identified vulnerability, it can propose mitigation strategies or best practices to enhance security.
- **Automatically generate performance testing scenarios based on system architecture**: By analyzing a system's architecture and its components, Generative AI can craft performance testing scenarios tailored to the system's unique structure and potential bottlenecks, ensuring that performance tests are both relevant and comprehensive.

## Deployment

- **Automated CI/CD pipeline creation and management**: With the integration of Generative AI, the process of setting up Continuous Integration and Continuous Deployment (CI/CD) pipelines can be automated. The AI can generate pipeline configurations based on the software stack, development practices, and target deployment environments, thereby ensuring streamlined integration, testing, and deployment of software changes.
- **Generate Kubernetes manifests, Makefiles, Dockerfiles, and Terraform Modules based on provided provisioning information**: Generative AI can be employed to autogenerate infrastructure as code (IaC) configurations, like Kubernetes manifests for container orchestration, Makefiles for build automation, Dockerfiles for containerization, and Terraform Modules for infrastructure provisioning. By analyzing the provided infrastructure requirements and system specifications, the AI produces optimized and tailored configurations, facilitating seamless deployment and scaling of applications.

## Operation and Maintenance

- **Deploy chatbots for code repository inquiries, technical support, and failure reporting**: Utilizing Generative AI, organizations can deploy sophisticated chatbots tailored to assist developers and users. These chatbots can answer inquiries related to code repositories, provide on-demand technical support, and facilitate the reporting of software failures. Their ability to learn from interactions allows them to continuously improve response accuracy and reduce manual intervention.
- **Scan clusters for diagnosing and triaging issues**: Generative AI can be instrumental in monitoring and scanning computational clusters or distributed systems. By analyzing system metrics, logs, and patterns, the AI can diagnose potential issues, rank them based on severity, and suggest triage steps. This proactive approach minimizes downtime and enhances system reliability.

## Support and Updates

- **Generate responses to Incidents and Customer feedback**: By leveraging Generative AI, support systems can automatically generate responses to incidents or customer feedback. This allows for immediate acknowledgment and initial troubleshooting, even if human intervention is later required for a more comprehensive solution. Rapid responses can enhance customer trust and satisfaction.
- **Analyze user feedback to automatically prioritize feature updates in future releases**: Generative AI can sift through and interpret large volumes of user feedback to identify patterns and trends. Based on this analysis, the AI can prioritize and suggest feature updates or bug fixes for subsequent releases. This data-driven approach ensures that development efforts align with user needs and preferences.

# POTENTIAL GENERATIVE AI TECHNIQUES FOR SOFTWARE DEVELOPMENT

Software development's increasing complexity necessitates innovative methodologies. Generative AI has emerged as a promising solution. This section critically evaluates its integration into software engineering.

We begin with Large Language Models (LLMs), detailing their training and fine-tuning processes for domain-specific applications. Next, the emphasis is placed on 'Prompt Engineering', a technique chosen for its efficiency. It encompasses varied approaches—zero-shot, few-shot, chain-of-thought, and self-consistency—to enhance interactions with AI.

The discourse then extends to Semantic Search with Word Embeddings, highlighting its potential for improved code comprehension. Tools and plugins developed for Generative AI in software engineering are also examined.

Concluding this section, we introduce LLM Agents and Code Assistants, suggesting a collaborative future between AI and developers.

Overall, this section offers a concise exploration of the nexus between Generative AI techniques and software development.

## Training Protocols and Operations in Large Language Models (LLMs)

Generative Artificial Intelligence (GenAI) primarily operates through two fundamental processes: inference and training. The inference phase, often termed the forward pass, is initiated post-training, facilitating users in eliciting responses such as text generation or question answering. Conversely, the training phase or backward pass is where the model assimilates data from one or multiple databases, employing optimization algorithms to finetune its parameters for optimal output. These core processes are visually represented in Figure 2.

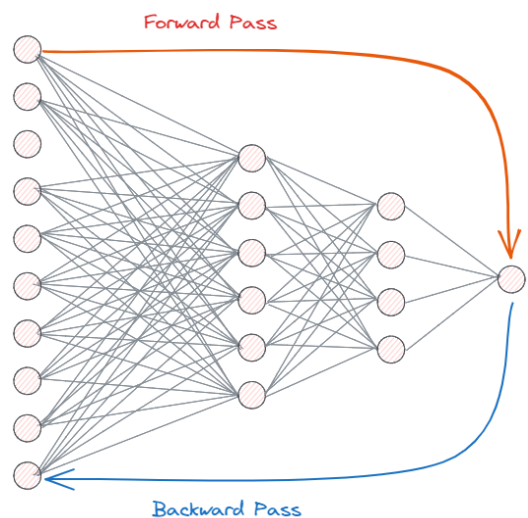

**Figure 2** - The two basic operations when using generative AI: Inference (forward pass) and training (backwardpass). Image adapted from [18].

Training stands as a pivotal element behind the recent advancements in the domain of Generative Artificial Intelligence (GenAI). At its core, training involves calibrating the parameters or weights of GenAI models, aligning them to generate desired outputs based on specific inputs. Such calibration demands vast datasets to ensure the model thoroughly comprehends the underlying problem. The calibration mechanism hinges on backpropagating the error discrepancy between the anticipated output and the model's generated output.

Our focus in this document is primarily on Large Language Models (LLMs). These models, a subset of Artificial Deep Neural Networks (DNN), build upon the Transformer architecture and excel in Natural Language Processing tasks [50]. Notably, for training open-source LLMs intended for code generation, The Stack dataset is a preferred choice. This dataset, with over 3 TB of data, encompasses code from 358 permissively licensed sources [38].

Historically, GenAI models were designed to represent a conditional probability distribution, denoted by: $P(X|Y = y)$, where $X$ signifies the observed data and $y$ represents the target [42]. This conveys that GenAI models encapsulate the probability structure of a problem, diverging from the traditional focus on class discrimination. To elucidate, GenAI models primarily train to synthesize data in alignment with a specific objective. For instance, a model might be trained with the aim of generating a cat image, subsequently producing assorted renditions of cat imagery, as visualized in Figure 3.

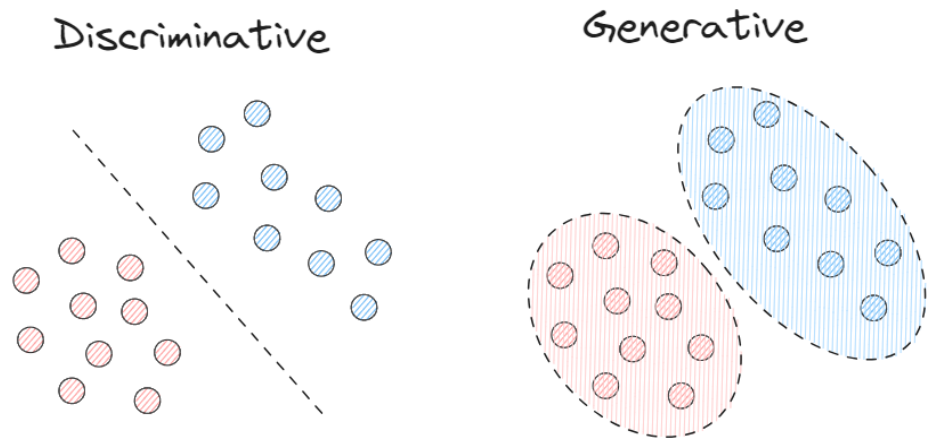

**Figure 3** - Example of how discriminative and generative models model a problem. Image adapted from [35].

For the case of LLMs, in general, these models can be trained to model three categories of problems [46]:
1. *Sequence-to-sequence (seq2seq):* This approach trains the model on a conditional probability distribution represented as $P(X|Y)$. Here, $Y$ could be an English sentence meant for translation or a posed question, while $X$ could signify its German translation or the respective answer. Essentially, the model ingests a word sequence and is tasked to generate a corresponding token sequence as output, a paradigm exemplified by the original Transformer model.
2. *Masked Language Modeling*: This strategy trains the model to discern the most probable word or token to be inserted into a text containing blanks. This methodology is apt for bidirectional models, as they have the advantage of contextual information from both before and after the omission.
3. *Next Token Prediction*: This approach focuses on equipping the model to anticipate the subsequent word that logically succeeds an input sequence. This method is tailored for unidirectional models, which rely solely on the preceding context when predicting the upcoming word.

Training models, especially in the domain of GenAI, can be broadly segmented into two primary categories:
1. **Supervised Learning**: This paradigm remains the gold standard for training both DNNs and LLMs. It utilizes a dataset consisting of paired inputs and their corresponding outputs. During the training phase, the model has simultaneous access to these inputs and their expected outputs. Variants of supervised learning include:
    a. **Complete Training from Scratch**: Starting with randomly initialized parameters or weights, the model undergoes training via backpropagation leveraging a

comprehensive dataset of inputs and their associated outputs. Such training is computationally intensive, demands considerable data, and is time-consuming.

   b. **Fine-tuning**: This approach begins with a pre-trained model, typically trained on a generic dataset. Fine-tuning, through backpropagation, refines this model using a task-specific dataset. Compared to training from scratch, fine-tuning is faster and requires less data. Various nuances of fine-tuning are elaborated in a dedicated section.
   c. **Distillation** [19]: Operating similarly to fine-tuning, distillation diverges in its use of synthetic data produced by a larger LLM trained on an extensive dataset.
   d. **Step-by-step Distillation** [37]: Mirroring distillation, this approach enhances training data with a sequence of intermediary steps to guide problem-solving.
2. **Reinforcement Learning from Human Feedback** [44]: Here, an LLM, initially trained via supervised learning, undergoes further refinement using human feedback. Human evaluators rank various prompts and their ideal outputs. A subsequent reward model utilizes these rankings to mete out rewards. The LLM is then retrained based on these rewards. Positive rewards are garnered when the LLM's outputs align with human-designated rankings, whereas deviations lead to negative rewards. This approach is pivotal in generating instruction-compliant LLMs while curtailing the potential for generating inappropriate content.

## Cost Implications of Training LLMs

Figuring out the exact cost of training an LLM can be tricky because there are many factors that can change depending on the model. But it is clear that training a model takes much more computer power than just using it. This not only means higher costs but also affects the environment.

| Model | Number of Parameters | Hardware | Total time on a single GPU |
|---|---|---|---|
| Transformer-base | 65 million | 8 NVIDIA P100 | 96 hours / 4 days |
| Transformer-big | 213 million | 8 NVIDIA P100 | 672 hours / 28 days |
| Helmet | NA | 3 NVIDIA GTX 1080 | 1008 hours / 42 days |
| BERT-base | 110 million | 64 NVIDIA Tesla V100 | 5068.8 hours / 211.2 days |
| GPT-2 | 1.542 billion | 32 TPUv3 | 5376 hours / 224 days |

**Table 2** - Computational cost and training time of some LLMs. Adapted from [49].

| Model | Number of Parameters | Amount of data used in training |
|---|---|---|
| GPT-3 | 175 billion | 570 gigabytes |
| BLOOM | 176 billion | 1.61 terabytes |
| Megatron-LM | 8.3 billion | 174 gigabytes |
| Megatron-Turing | 530 billion | Not disclosed |

| Model | Number of Parameters | Amount of data used in training |
|---|---|---|
| BERT-large | 340 million | 3.3 billion words |

**Table 3** - Amount of data used for training some LLMs

In 2019, a study found that training a basic BERT model, which has 110 million parameters, costs between US$3,500 and US$12,500. This also has an environmental impact of about 0.65 tons of $CO_2$ [49]. Another study looked at the Sabiá model, which has between 7 and 65 billion parameters and is based on the LLaMA model. The cost for this was US$9,000 to US$80,000 [47]. To understand this on a bigger scale, OpenAI's GPT-3 model, with 175 billion parameters, used 190,000 kWh of energy. This is the same amount of energy as driving a car to the moon and back.

The cost of training a model can depend on the type of model, the optimization method, the batch size, and the sequence length. These factors can change how much memory and processing power is needed. For example, Table 2 shows the computer needs and training times for some LLM models. At the same time, Table 3 lists the amount of data used to train these models.

Because of all this, organizations that want to use generative AI should understand that they will need strong computer systems to do their research and testing.

# Fine-tuning LLMs

Although often classified as a subset of training, fine-tuning warrants a dedicated section in this whitepaper due to its widespread application in LLMs. The essence of fine-tuning lies in its ability to customize and refine a pre-trained model for specific tasks, bolstering its accuracy and performance [14]. To understand this in the context of LLMs, one must first recognize that during the initial training phase, a model is exposed to vast amounts of generic data. This enables it to grasp broad language patterns and features, culminating in a pre-trained model.

However, the breadth of this general training often means the resulting model is not intrinsically optimized for specific tasks, datasets, or nuanced behaviors. Herein lies the significance of fine-tuning: it rectifies these gaps. A testament to its popularity, fine-tuning reduces computational complexity compared to initiating training from scratch while retaining commendable customization potential [10]. As depicted in Figure 4, fine-tuning emerges as the method of choice for those seeking significant model customization without the constraints of high-end hardware requirements. Conversely, as Figure 4 also indicates, while Prompt Engineering alleviates the need for cutting-edge hardware, it falls short in offering deep customization capabilities.

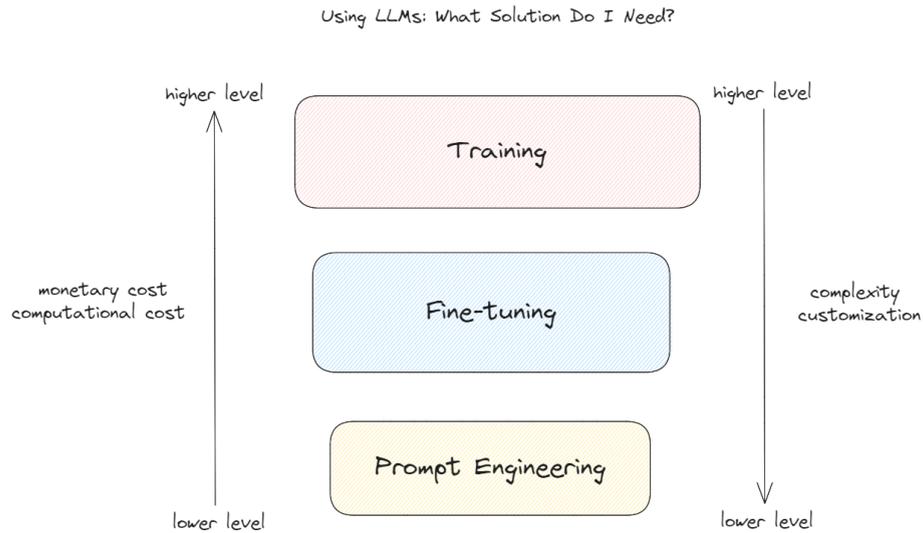

**Figure 4** - Levels of complexity and customization between training from scratch, fine-tuning a pre-trained model and prompt engineering.

Figure 5 delineates the extent of customization achievable through fine-tuning an LLM. In essence, fine-tuning equips a pre-trained LLM with abilities spanning from language translation and question-answering to sentiment analysis, comprehensive text summarization, language trick detection, and the verification of information within textual content.

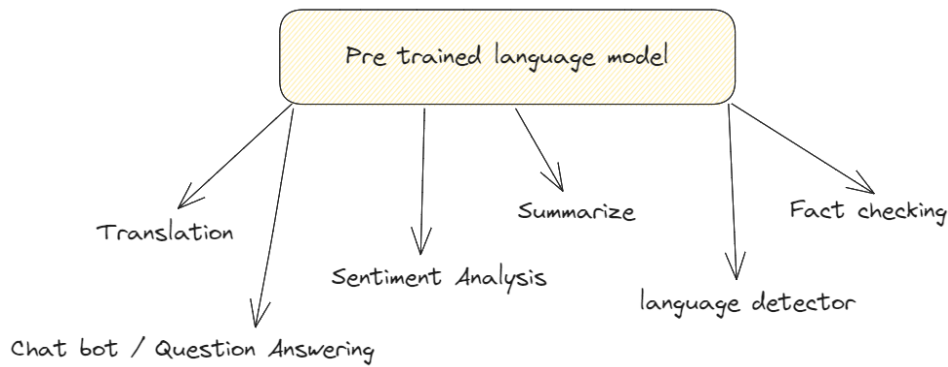

**Figure 5** - For which tasks a pre-trained language model can be fine-tuned for.

Much like the multifaceted nature of training from scratch, fine-tuning also encompasses a variety of methods. Some of the most prominent ones include:

1. **Conventional Fine-Tuning**: This process involves refining a pre-trained model on a specific dataset tailored for a particular task. By updating the model parameters with this new dataset, the model becomes better attuned to the nuances of the targeted task.
2. **Transfer Learning:** Here, a model that has been pre-trained on one task is repurposed as a foundation for a related but distinct task. This method capitalizes on the knowledge acquired during the initial pre-training, facilitating quicker model training and often resulting in enhanced performance for the new task.

3. **Parameter-Efficient Fine-Tuning**: This technique optimizes the fine-tuning process by selectively updating only a subset of the model's parameters. The objective is to achieve effective model refinement without the computational burden of tuning all parameters.
4. **Learning from Textual Interactions**: In this approach, the model is trained using pairs of texts, wherein one serves as the context and the other as the response. This training methodology hones the model's capability to generate contextually relevant responses, proving invaluable in applications like conversational AI or Q&A systems.

## Prompt Engineering

The *Prompt Engineering* strategy focuses on designing and refining prompts to boost the performance of language models. This method includes creating clear instructions or specific questions that guide or shape the responses of the models. The goal is to produce answers that are clear, correct, and aligned with what the user wants. Much of our understanding comes from the Prompt Engineering Guide [22] website.

### Zero-Shot Prompting.

This method involves giving a task to a language model without offering examples of how to do it (see Figure 6). Thanks to the vast data the model was trained on, this approach can be used for some tasks without any issues [21, 52]. In simpler terms, zero-shot prompting works well for basic tasks.

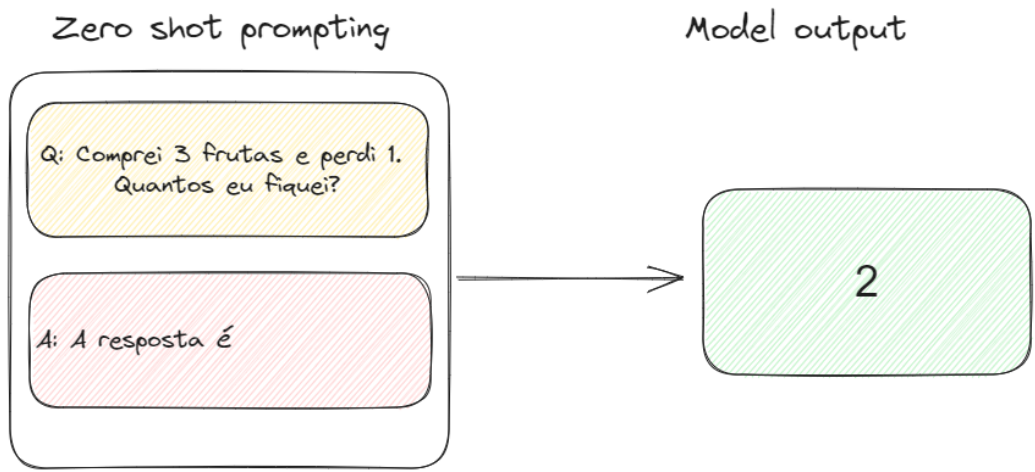

**Figure 6** - Diagram of how the "zero-shot prompting" technique works.

### Few-Shot Prompting

The few-shot prompting method gives both instructions and examples to a language model for a specific task (see Figure 7). This method helps the LLM make responses in the right format and guides its thinking, helping it learn in context. It is especially useful when we want the model's output in a special format that is hard to explain. Similarly, there is the One-Shot Prompting technique. Here, we provide the model with just one example instead of multiple ones.

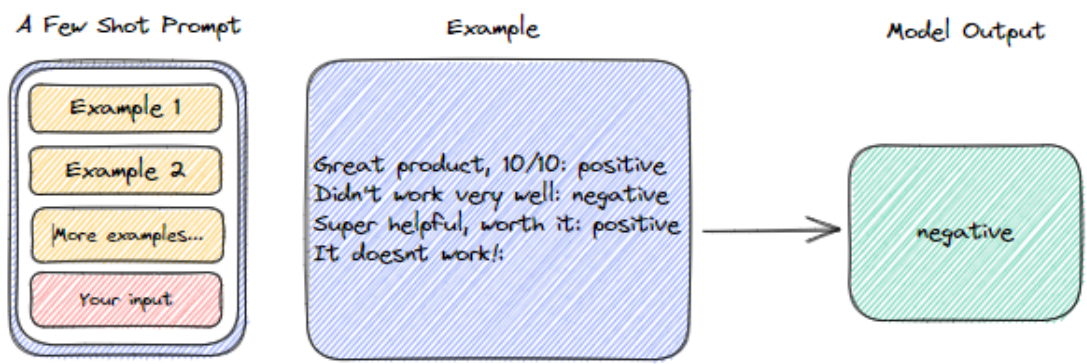

**Figure 7** - Diagram of how the "Few shot prompting" technique works. Image taken from [8, 20, 25].

## Chain-of-Thought Prompting

The Chain of Thoughts (CoT) technique involves a series of steps that help large language models (LLMs) reason better [7, 24]. Normally, LLMs struggle with tasks that need many steps or a lot of reasoning. But with CoT, they're encouraged to think step by step, which helps them solve problems and come up with better answers. As a result, LLMs perform better on tests of common sense, math, and other types of reasoning. It also makes sure that the model's responses are clear and stick to the topic, so the answers are more detailed and make sense (see Figure 8).

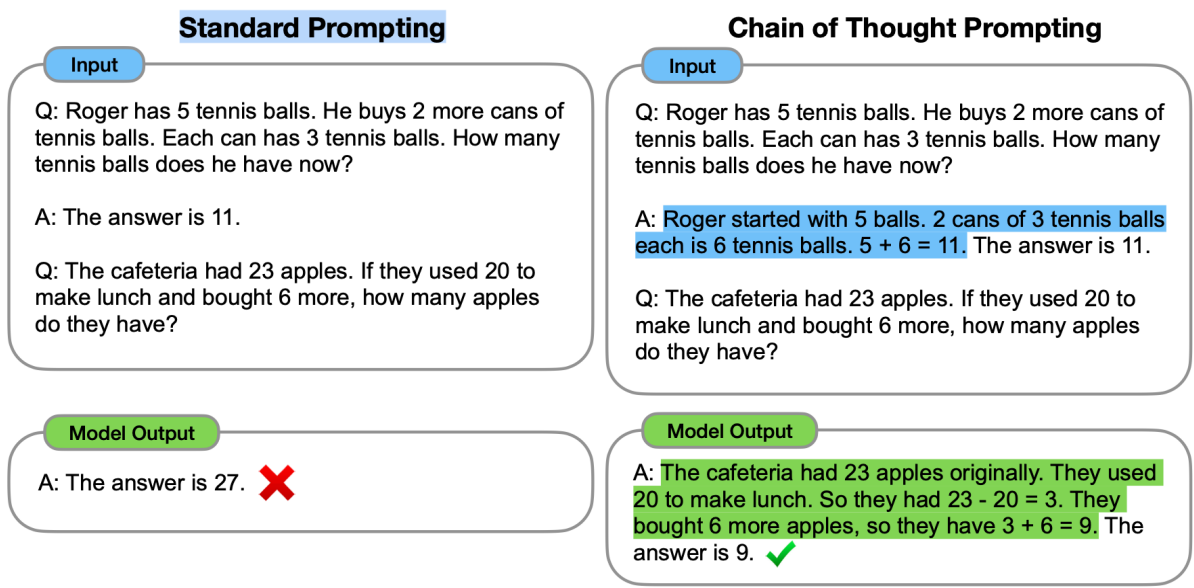

**Figure 8** - Comparing the traditional prompt with the Chain of Thought prompt. Image from [7].

There are two versions of the CoT technique: Zero-Shot-CoT and Few-Shot-CoT. With Zero-Shot-CoT, the language model is told to think things through step by step, without any examples (see Figure 9). On the other hand, Few-Shot-CoT gives the model a few examples to guide its thinking for a specific task.

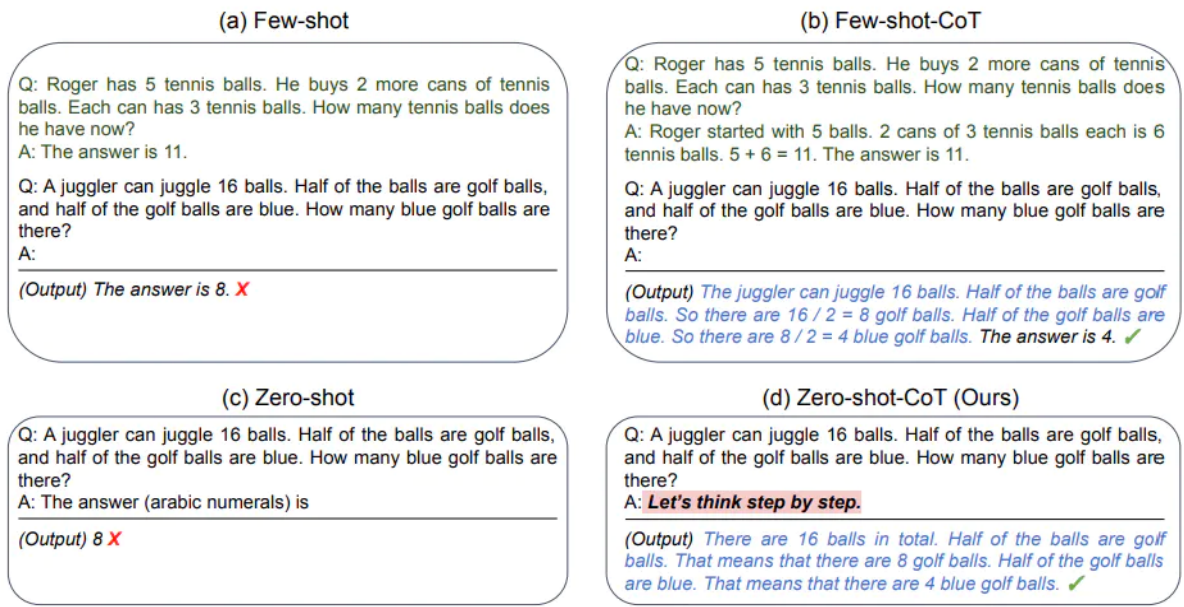

**Figure 9** - Difference between Few-shot and Zero-shot with and without the use of CoT. Image taken from [53].

## Self-Consistency

Self-consistency is a decoding technique that aims to improve how pre-trained language models reason. Instead of the model taking the first, most obvious reasoning path, this technique explores various reasoning paths. Then, it picks the answer that shows up most consistently across these paths (see Figure 10. This approach helps in situations where there are many ways to come to a correct answer, like in common reasoning tasks or arithmetic.

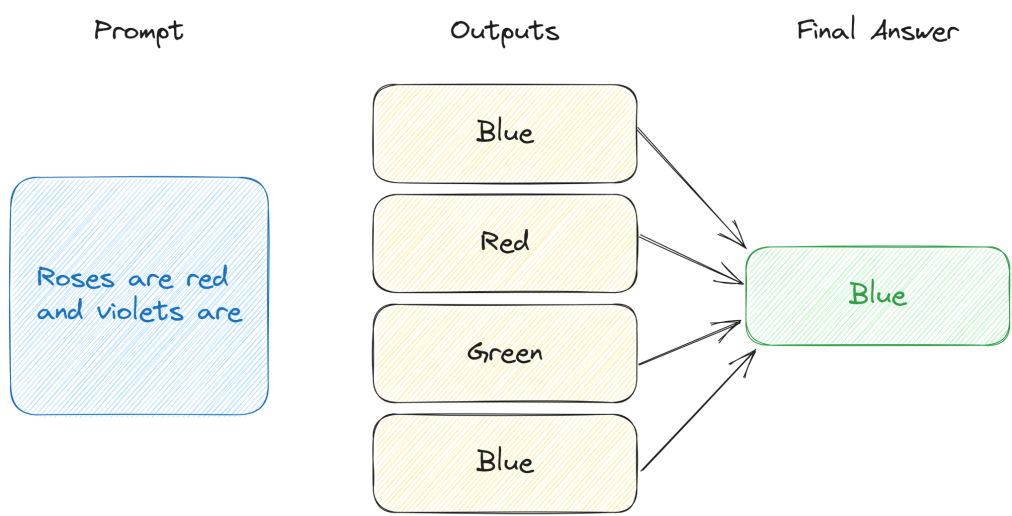

**Figure 10** - Diagram showing the Self-consistency method. Image from [9].

Here is how it works: First, the model is given a question or task. Instead of taking the most common path, which might be incorrect, it explores different reasoning paths using the Chain of Thoughts (CoT) technique. This results in a series of sentences that show how the model is thinking.

After this, the Self-Consistency method picks the answer that is most consistent across these reasoning paths. This answer is taken as the final response of the model, as illustrated in Figure 11.

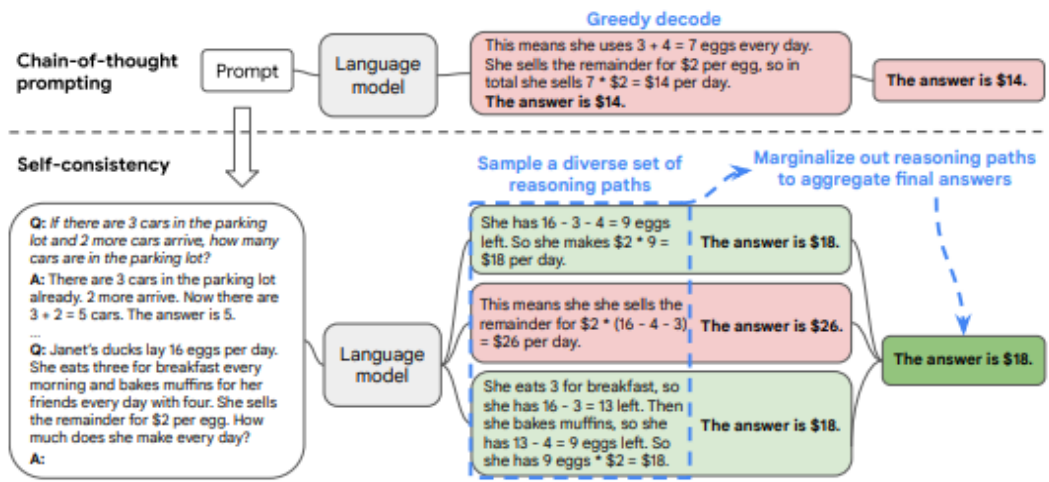

**Figure 11** - Self-consistency example. Image taken from [51].

## Generate Knowledge Prompting

Large Language Models (LLMs) acquire vast amounts of information during their pre-training phase. However, to enhance their performance, it can be beneficial to supplement your prompts with additional relevant information [29, 36]. Providing this extra context can address challenges, such as the model producing inaccurate information or "hallucinations."

Essentially, the core principle of this technique is to prompt the model to recall or generate pertinent knowledge about a given topic before producing the final output. You have the option to craft a single prompt that both recalls the information and completes the task or use two separate prompts: the first to draw out the knowledge and the second to act based on the information from the initial prompt (see Figure 12).

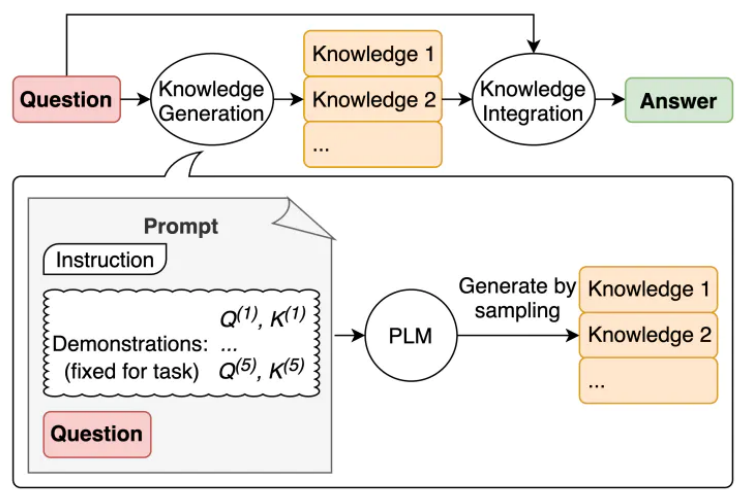

**Figure 12** - Representation of Generate Knowledge Prompting. Image taken from [41].

## Tree of Thoughts

The Tree of Thoughts (ToT) framework guides a model to think in a structured manner by breaking down thoughts into intermediate steps, aiding in general problem-solving [54]. This approach is especially beneficial for tackling complex tasks. The underlying principle of ToT is the recognition that thoughts and ideas are not strictly linear. Instead, like branches on a tree, they stem from a core concept and spread out into various related yet distinct directions (see Figure 13). An LLM traverses this "tree" to derive a thorough and coherent response that encapsulates the essence of the initial, broad question.

To succinctly differentiate between CoT and ToT: ToT facilitates the exploration of coherent segments of text or "thoughts" that act as stepping stones towards problem-solving. Each branching point or node serves as a partial resolution to the presented problem, building upon the given input and the cumulative sequence of thoughts. While it shares the "step-by-step thinking" foundation with CoT, ToT emphasizes meticulous assessment at each step or intermediate thought.

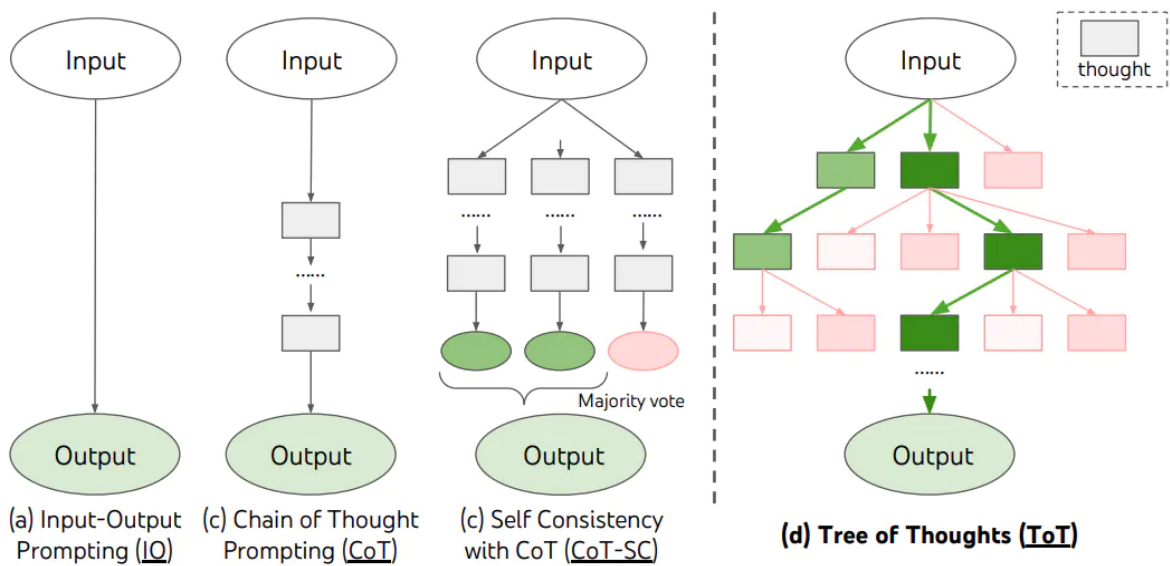

**Figure 13** - Comparison of techniques with ToT. Image taken from [26].

Example: Imagine you are trying to solve a math problem. Using the CoT approach, you would tackle the problem step by step, like following a straight path. For instance, if the problem is to solve $2x + 5 = 11$, CoT would go: "First, I will subtract 5 from both sides. Now, I have $2x = 6$. Next, I'll divide both sides by 2. So, $x = 3$."

On the other hand, using the ToT approach, you might start with the same problem but think of multiple ways to solve it, branching out like a tree. So, one would think, "I could use algebra, or I could plug in numbers to see which one works." If you go the algebra route, one would then think, "Do I subtract 5 first or do something else?" Each of these decision points is like a branching node on a tree, leading you to the final answer, but allowing for exploration along the way.

In essence, while both methods aim to solve the problem, ToT lets one explore and evaluate various routes before settling on an answer.

## Automatic Reasoning and Tool-use

The Automatic Reasoning and Tool Use (ART) approach combines the Chain-of-Thought method with various tools to achieve tasks [27]. Here's how ART operates:

When presented with a new task, ART selects appropriate reasoning examples and tools from its library. If the task execution requires an external tool, ART temporarily halts result generation. Once the tool provides the needed result, ART incorporates this information into the final output. This approach enables the model to learn from examples and apply the correct tools even if it hasn't been specifically trained for that task. Moreover, ART has a feature that allows human input (see Figure 14). This means users can provide feedback to refine the process or introduce new tools and libraries when necessary.

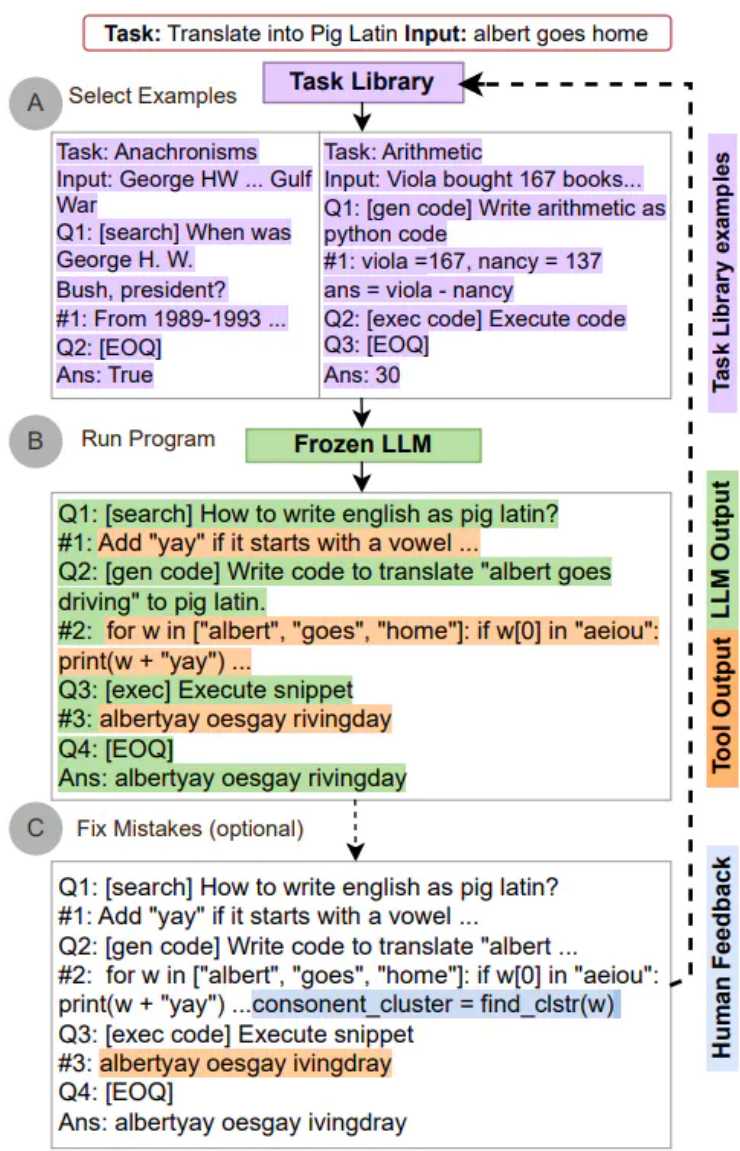

**Figure 14** - ART functioning pipeline. Image taken from [45].

## Automatic Prompt Engineer

The Automatic Prompt Engineer (APE) is designed to automatically create guidelines for specific tasks [57]. It works by taking in income statements and then offering multiple statement options (see Figure 15). These options are either generated directly or through a step-by-step process that considers how similar the meanings of the statements are [23]. After generating these instructions, the model carries them out. The best option is then picked based on performance evaluations. APE is particularly useful when it is hard for humans to determine the right instruction for a task, or when there is a long list of tasks to complete.

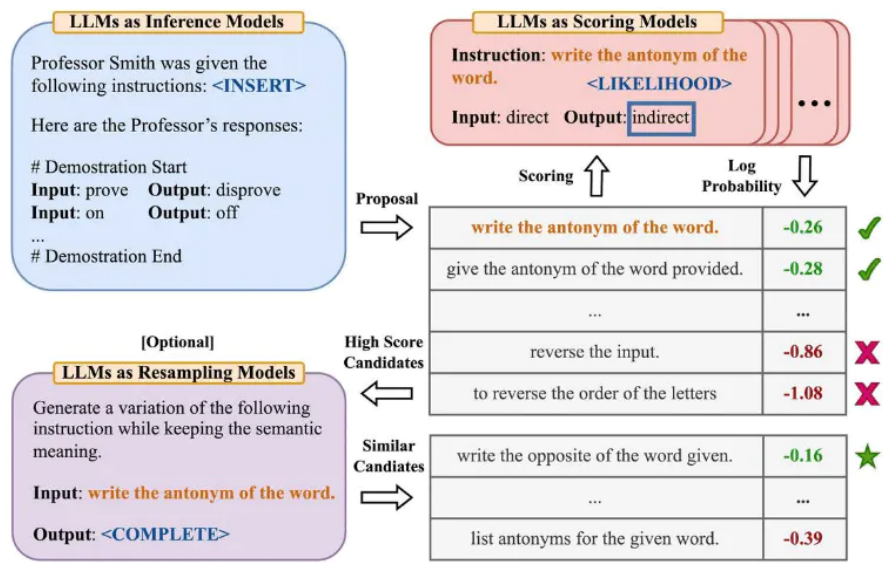

**Figure 15** - APE flow. Image taken from [57].

## Active-Prompt

The Active-Prompt methodology is an extension of the CoT approach, which primarily depends on a predetermined set of human-annotated examples [32]. Active-Prompt refines this by adapting Language Models (LLMs) to task-specific example prompts that integrate human-crafted CoT reasoning (see Figure 16). The procedure for Active-Prompt is outlined below:

1. Consult the LLM, with or without certain CoT examples.
2. Generate several possible responses (denoted as k answers) for a collection of training questions.
3. Compute an uncertainty metric derived from the k responses.
4. Choose questions exhibiting the highest uncertainty to be annotated by human experts.
5. Use the newly annotated instances to deduce the answers to each question.

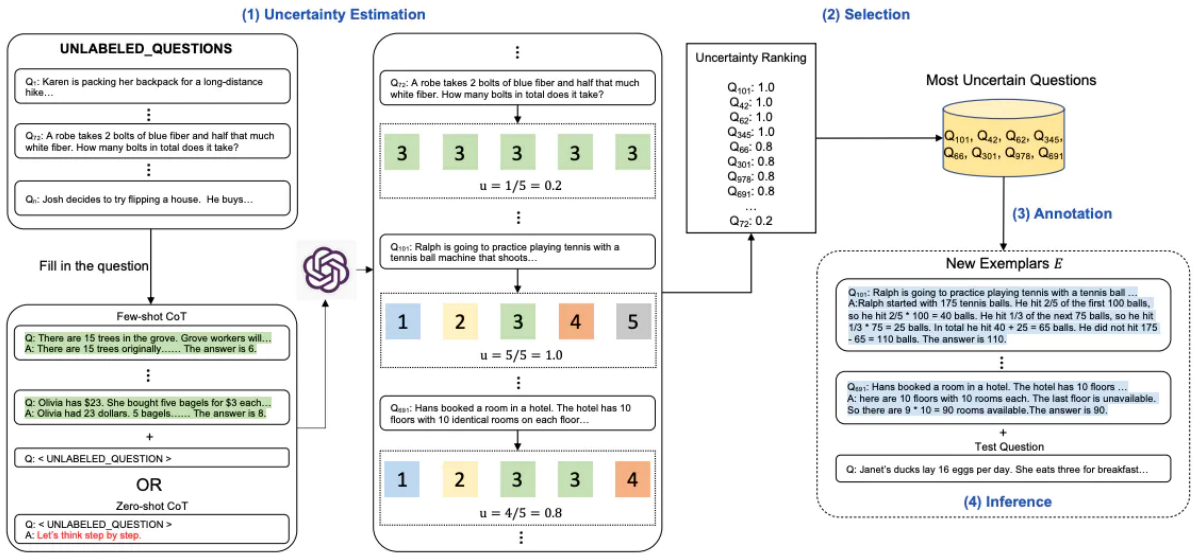

**Figure 16** - Active-Prompt flow. Image taken from [33].

In essence, Active-Prompt enhances CoT methods by harnessing the capabilities of LLMs and tailoring example prompts, thereby addressing the challenge of selecting optimal examples for diverse tasks.

## Directional Stimulus Prompting

Directional Stimulus Prompting is a method designed to steer an LLM towards producing a specific type of answer [28]. By offering a hint, this method focuses the LLM's attention on certain aspects when crafting its response (see Figure 17). One common application of this technique is in creating summaries.

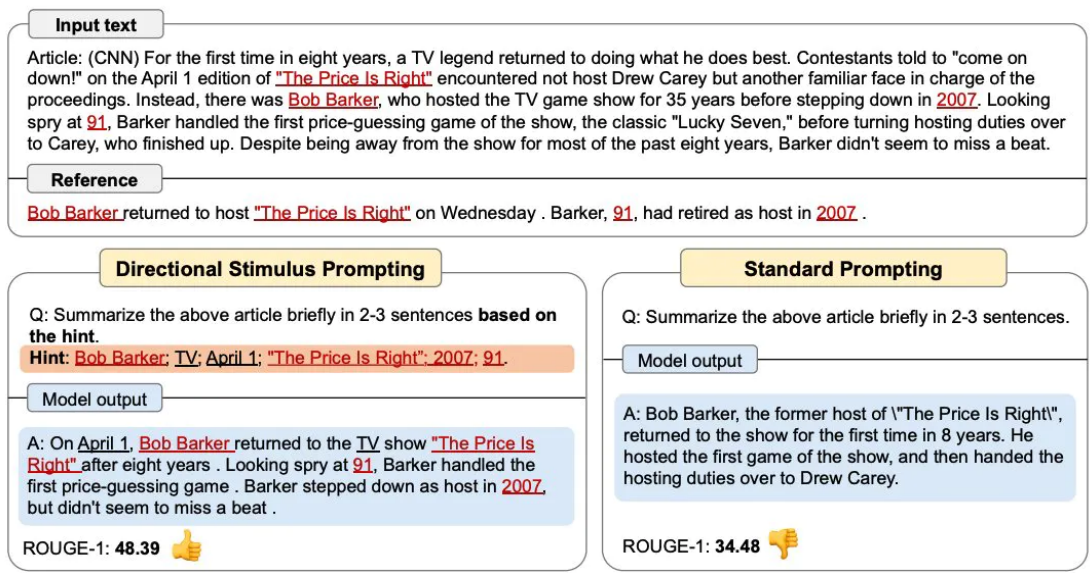

**Figure 17** - Example of operation. Image taken from [39].

## ReAct

The ReAct framework is designed to enhance the decision-making capabilities of LLMs by integrating them with external tools. These tools provide extra information, ensuring answers that are more accurate and factual [55]. In this approach, LLMs produce verbal reasoning paths and actions tailored for specific tasks (see Figure 18). This setup facilitates dynamic reasoning, helping the system to devise, uphold, and modify action plans. This process can be thought of as a combination of Reasoning (RE) and Acting [31]. Notably, LangChain incorporates the ReAct framework by default, allowing the creation of agents that execute tasks.

By interacting with a straightforward Wikipedia API, ReAct effectively addresses common issues like hallucination and the spread of errors found in CoT reasoning. The outcome is the generation of human-like task-solving paths that are clearer to understand compared to standard methods without such reasoning paths [15].

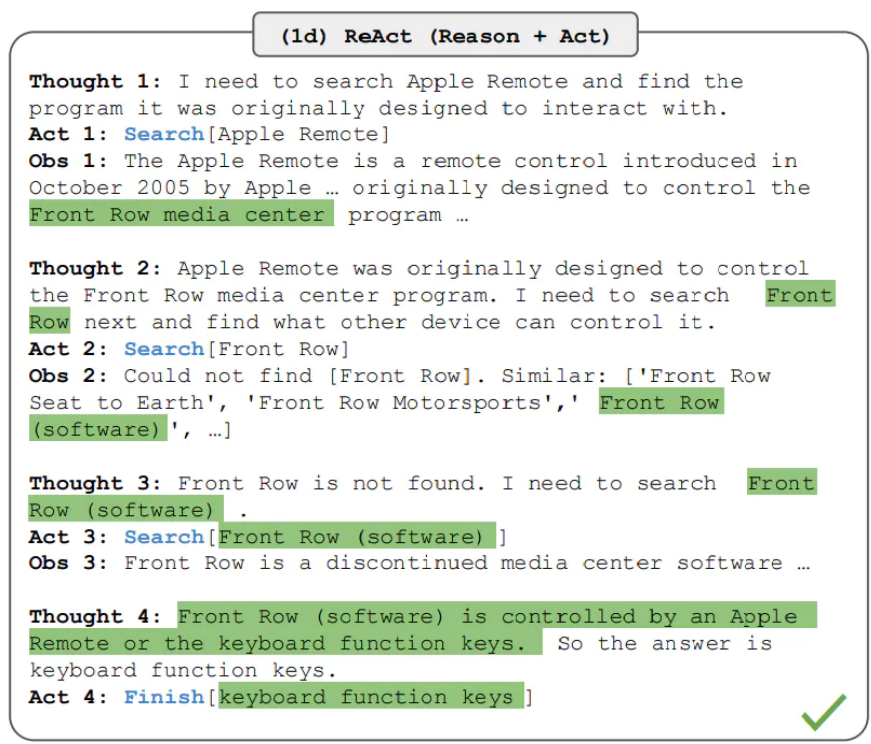

Figure 18 - Representation of ReAct. Image taken from [55]

## Multimodal Chain-of-Thought

The Multimodal Chain-of-Thought (Multi CoT) differs from the Chain-of-Thought (CoT) in that it combines both vision and text, rather than just focusing on language [56]. This approach allows a model to break down a complex problem into a set of smaller steps using inputs from different sources (see Figure 19). It then uses these steps to determine the answer [30]. The process is divided into two main stages. In the first stage, the model creates reasoning using information from multiple sources. In the second stage, it uses this reasoning to come up with the final answer.

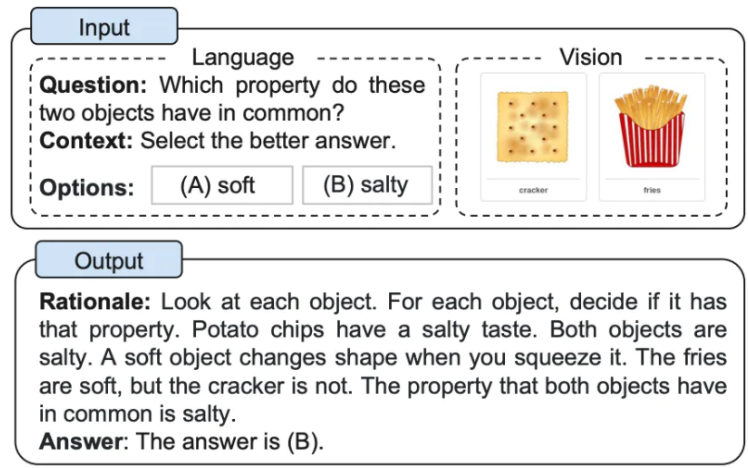

**Figure 19** - Representation of the Multimodel CoT. Image taken from [56].

## Semantic Search

Semantic search is a technique used in search engines that focuses on understanding the deeper meaning of user queries and the context of information on the web [13]. Instead of just looking for matching keywords like traditional methods (Keyword Search), semantic search delves into what the user truly means with their query and aims to give more accurate results. It looks at the relationship, context, and significance of the terms in the search. By using advanced tools such as natural language processing, ontologies, and machine learning, semantic search can break down and interpret both the user's question and the content of web documents. This way, it can recognize things like related words, opposite words, and broader or narrower terms, and even make connections between different ideas.

Some key concepts for understanding how semantic search works in the context of generative AI are: vector embeddings and context.

### Vector Embeddings

In the realm of natural language processing, a Vector Embedding is essentially a way to represent a word as a vector in a space with many dimensions. This representation is designed so that words with similar meanings or uses are positioned closely together in this space, which helps to capture their context and semantics.

To create these vectors, machine learning techniques, such as neural networks, are utilized. These methods are trained on vast amounts of text data [17]. As they train, the algorithms learn to assign vectors to words in such a way that words used in similar ways or with related meanings have vectors that lie near one another [13]. The act of turning a data item into a vector is known as vectorization (see Figure 20).

Common models for training and using word arrays include word2vec (Google), GloVe (Stanford), ELMo (Allen Institute/University of Washington), BERT (Google), and fastText (Facebook).

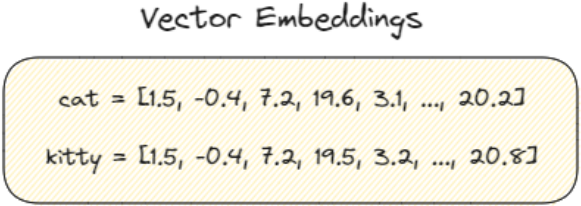

Figure 20 - Example of two word and their corresponding representation as vector embeddings

### Context

In models like GPT-3, which are considered large language models, the "context" denotes the text or data fed into the model as input [12]. This context is crucial as it helps the model generate text that is both coherent and relevant.

Whenever such a model is presented with a task or a question, it utilizes the provided context to craft a suitable reply. This context is not just limited to the immediate question or directive; it also encompasses any past interactions, as well as any texts, documents, and other relevant information. It is essential to provide the model with adequate context, especially since the model might be relying on older information or might not be aware of specific, crucial details needed for an accurate response.

### In-context Learning

In-context learning is a technique that allows language models to perform tasks based on prompt information [1]. There are two approaches that are being widely used within in-context learning [11, 40].

In-context learning involves showing the model a set of examples that display the desired input and output (see Figure 21). After this, a new test input is introduced, and the model is tasked with predicting an outcome based on the knowledge it acquired from the initial examples.

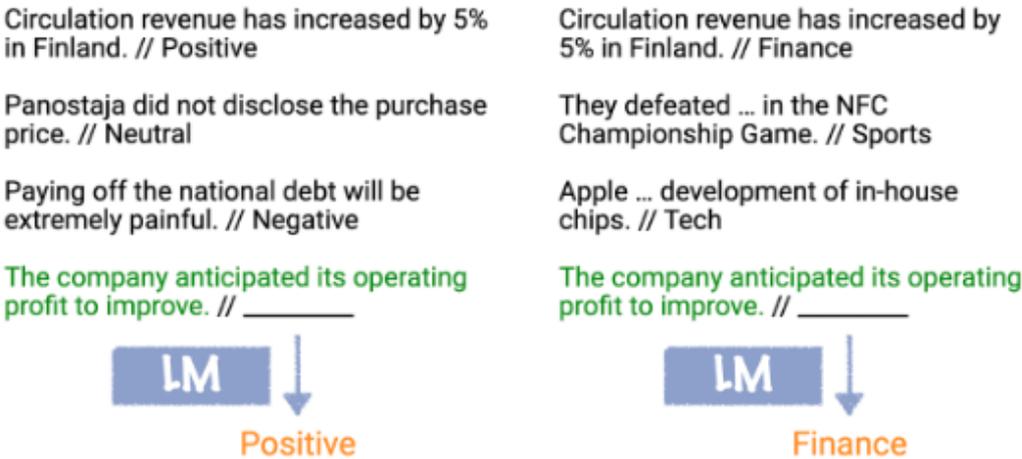

**Figure 21** - Representation of In-context learning through examples

For the model to accurately address new test inputs, it must apply what it learned from the examples. This includes understanding the nature of the news, determining if it is positive or negative, and knowing how to categorize it. The model should also grasp details about the proper formatting.

### In-context learning through new data sources

The model can receive context from a wide variety of data sources, including books, personal documents, datasets, legal papers, project technical documentation, newspapers, and notes, among others. Introducing these data sources paves the way for many new applications of LLMs. It lets us input both personal and corporate data and even supply the model with real-time information.

However, a significant challenge in using these techniques is the need to continuously provide this context through the prompt in every session, as the model does not truly train on this data. Moreover, if your contextual documents are extensive, the number of tokens in each prompt can grow significantly, leading to increased usage costs

## Vector databases as context sources for LLMs

Using extensive textual sources as context for LLMs can be achieved by breaking down the text data into more manageable segments. A tokenizer can then transform these smaller text segments into tokens [13]. Once tokenized, an LLM embeddings API can turn each text segment into a series of arrays or vectors. These vectors can subsequently be stored in a vector database for future use.

It is important to note that the initial conversion of text to vectors through an embeddings API can be costly. Often, there are charges on a per-token basis when using such APIs. However, the silver lining is that this conversion process is a one-time activity for any given dataset. Once in the vector database, these vectors remain available for use across various sessions, prompts, and even different LLMs.

The next hurdle is retrieving the contextual data from the vector database to use in an LLM prompt. We can explore two potential strategies to achieve this.

### Querying the Vector Database Prior to LLM Prompting

One method involves generating a vector representation of the user or application query, which is then sent to the vector database index for a semantic search. After identifying the top results, the original text data from these results is appended to the prompt. This enriched prompt is then processed by the LLM.

### Empowering LLM to Initiate Vector Database Queries

A more advanced approach lets the LLM determine when it requires additional information and initiate its own vector database queries based on the received prompt. This would, in essence, allow the LLM to identify and retrieve the information it deems necessary. For instance, if the prompt suggests "if you require additional information, formulate a natural language query to search the vector database," the LLM could generate context-relevant queries. These queries can be processed using frameworks that interface with the vector database, extracting the relevant results. The information is then fed back into the LLM's prompt.

The primary objective of both methods is to supply the LLM with concise data for each prompt, promoting cost-effective utilization. By focusing on delivering only the most pertinent context via semantic search, the quality of the information provided to the LLM is enhanced, mitigating the likelihood of generating inconsistent responses.

## Plugins and Tools for Large Language Models

The vast potential of the Large Language Models can be further harnessed and directed using Plugins and Tools. These specialized components are designed to augment the intrinsic capabilities of LLMs, offering enhanced functionality, precision, and usability in various applications.

1. **Task-specific Fine-tuning**: Some plugins facilitate fine-tuning of LLMs for specific tasks, such as sentiment analysis, code generation, or translation. This ensures better performance as the model becomes more attuned to the nuances of the desired task.
2. **Interactivity Enhancements**: Certain tools can enhance the interactivity of LLMs, providing real-time feedback, aiding in iterative model training, or assisting in dynamic prompt generation to retrieve more accurate responses.
3. **Visualization Tools**: Understanding the inner workings and decision-making processes of LLMs can be challenging. Visualization plugins can offer insights into attention mechanisms, layer activations, and other internal aspects, promoting better model interpretability.
4. **Optimized Inference Engines**: LLMs can be resource-intensive. Plugins that provide optimized inference capabilities can speed up model predictions, making them more feasible for real-time applications or deployment on resource-constrained platforms.
5. **External API Integration**: Plugins can equip LLMs with the capability to access and integrate data from external APIs. This allows the model to pull real-time information, enriching its responses or making decisions based on up-to-date data. For instance, an LLM can be augmented to fetch recent news articles, check the weather, or interact with third-party data sources to generate contextually relevant and informed outputs.

In conclusion, the integration of plugins and tools with LLMs marks a promising avenue for refining and expanding the practical applications of these models. By leveraging these additional components, developers and researchers can maximize the benefits of LLMs, ensuring that they remain at the forefront of the AI landscape.

# LLM Agents

LLM Agents are sophisticated software entities designed to automate the execution of tasks. Equipped with access to a comprehensive toolkit or set of resources, these agents intelligently determine the best tool or method to use based on the specific input they receive. The operation of an LLM agent can be visualized as a dynamic sequence of steps, meticulously orchestrated to fulfill the given task. Significantly, these agents have the capability to use the output from one tool as input for another, creating a cascading effect of interlinked operations.

## BabyAGI

BabyAGI is an advanced task management system, powered by OpenAI's cutting-edge artificial intelligence capabilities [43]. In tandem with vector databases like Chroma or Weaviate, it excels in managing, prioritizing, and executing tasks with high efficiency. Using OpenAI's state-of-the-art natural language processing, BabyAGI can formulate new tasks aligned with specific objectives. Moreover, it boasts integrated database access, enabling it to store, recall, and utilize information that is pertinent to the context of its tasks. Essentially, BabyAGI represents a streamlined version of the Task-Driven Autonomous Agent, incorporating notable features from platforms like GPT-4, Pinecode vector search, and the LangChain framework to independently craft and execute tasks.

The operational flow of BabyAGI can be broken down into the following four steps (see Figure 22):

1. Extract the foremost task from the pending task list.
2. Relay the task to the dedicated execution agent for processing.
3. Refine the derived result, subsequently storing it in either Chroma or Weaviate for future reference.
4. Formulate new tasks, dynamically adjusting the priority of the task list based on both the overarching objective and the outcomes of previously executed tasks.

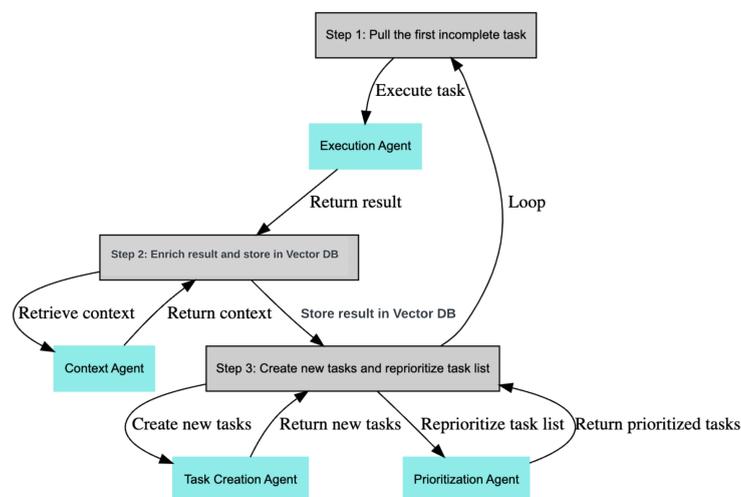

**Figure 22** - A visual depiction illustrating the operational flow of BabyAGI. Image sourced from [43]

### AgentGPT

AgentGPT is a robust platform tailored for the creation and deployment of autonomous AI agents. Once a particular objective is defined for these agents, they delve into a relentless loop of task generation and execution, striving to meet the stipulated goal.

At the heart of its operation lies a chain of interconnected language models (or agents). These models collaboratively brainstorm the optimal tasks to meet an objective, execute them, critically assess their performance, and iteratively devise subsequent tasks. This recursive approach ensures that AgentGPT remains adaptive, learning, and refining its strategies with each loop to inch closer to the objective.

## Code Assistants: Enhancing IDEs with AI-driven Plugins, Extensions, and Add-ons

Code assistants are advanced tools designed to assist developers in the code-writing process. They often are implemented as IDE plugins, extensions, or add-ons, and are capable of suggesting code completions, identify and rectify bugs, provide optimization recommendations, and simplify recurring coding tasks. Incorporating Generative AI models, they analyze coding patterns and furnish insights that streamline the development workflow, thereby accelerating code generation and elevating the quality of output.

### GitHub Copilot

Developed through a collaboration between GitHub and OpenAI, GitHub Copilot harnesses the capabilities of the Codex generative model, aiding developers in writing code more efficiently [5]. Described as an AI-powered programming companion, it presents auto-complete suggestions during code development. GitHub Copilot keenly discerns the context of the active file and its related documents, proposing suggestions directly within the text editor. It boasts proficiency across all languages represented in public repositories.

Copilot X builds upon the foundation of Copilot, offering an enriched experience with chat and terminal interfaces, enhanced support for pull requests, and leveraging OpenAI's GPT-4 model [6]. Both Copilot and Copilot X are compatible with Visual Studio, Visual Studio Code, Neovim, and the entire JetBrains software suite.

### AWS CodeWhisperer

Amazon CodeWhisperer is a machine learning-driven code generator that proffers real-time coding recommendations [2]. As developers script, it proactively presents suggestions influenced by the ongoing code. These propositions range from concise comments to elaborately structured functions. Currently, CodeWhisperer is attuned to a multitude of programming languages, including Java, Python, JavaScript, TypeScript, and many more. The tool seamlessly integrates with platforms such as Amazon Sagemaker Studio, JupyterLab, Visual Studio Code, JetBrains, AWS Cloud9, and AWS Lambda.

### Bard to Code

Bard is often categorized as conversational AI or chatbot. Owing to its extensive training on a myriad of textual data, Bard demonstrates an adeptness in producing humanesque textual responses to a diverse spectrum of prompts. Moreover, it possesses the dexterity to produce code across various programming languages, including but not limited to Python, Java, C++, and JavaScript [3].

### CodeGPT

CodeGPT, tailored for developers, excels in providing code suggestions, rectifying errors, showcasing examples, and answering queries related to programming [4]. Serving as an indispensable programming assistant, CodeGPT accelerates the development trajectory and amplifies developer efficiency. Thanks to its vast training on an array of source codes and related data, CodeGPT can craft code, elucidate programming concepts, assist in debugging, and even contribute to technical documentation. It exhibits versatility across languages such as Python, JavaScript, Java, C++, and more. Notably, CodeGPT is available as a vscode extension.

### Tabnine

Tabnine epitomizes the AI-driven code assistant, propelling development speed with instantaneous code completions across various languages and IDEs [16]. As an AI co-pilot, it presents invaluable features such as code snippets, predictions, suggestions, and hints, profoundly influencing coding speed. Tabnine is committed to maintaining the integrity of source code; it solely utilizes open-source code with permissive licenses, such as MIT, Apache 2.0, and BSD variants, for training its Public Code AI model. Irrespective of the subscription tier, user code and AI data are strictly preserved and not utilized to train any models outside of private code models.

# Conclusion

This whitepaper explored the promising intersection between Generative AI and Software Development. We began by outlining the key stages and artifacts involved in the software development lifecycle, providing context on the process that Generative AI aims to enhance.

Next, we identified potential demands within software engineering that Generative AI is poised to address. Numerous applications were discussed, spanning project management, requirements gathering, system architecture, database design, coding, testing, deployment, operations, and user support.

We then delved into the AI techniques relevant to this domain. This included training protocols, fine-tuning approaches, prompt engineering strategies, semantic search leveraging vector embeddings, and tools like plugins and code assistants that augment LLMs.

Our analysis found that Generative AI, when thoughtfully implemented, can bring formidable productivity and quality improvements to software development. Automating repetitive tasks, generating artifacts and prototypes, optimizing queries and code, identifying vulnerabilities pre-emptively, and providing intelligent assistance are some of the key benefits.

However, simply integrating the latest AI models into the development process without a clear methodology often yields suboptimal results. Organizations must invest time into prompt engineering, specialized fine-tuning, UX design, and seamless human-AI collaboration. Striking the right synergy between generative capabilities and human oversight is pivotal.

As Generative AI continues maturing, its integration with software engineering will only deepen. We hope this whitepaper stimulates further research and inspires new applications at the intersection of these domains. The future appears bright for developers leveraging AI as an indispensable partner in building impactful software.

# About the Author(s)

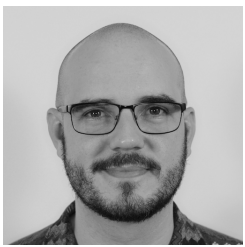

**Filipe Calegario** is an associate professor at the Center for Informatics of UFPE. He was an industrial researcher at the SENAI Institute of Innovation for Information and Communication Technologies. He holds a Ph.D. in Computer Science from the Center for Informatics (CIn), UFPE, with a research focus on electronic prototyping and digital manufacturing in the design of new interfaces for musical expression. In 2015, he conducted a sandwich Ph.D. program in the field of musical technology at McGill University, Canada. He is one of the founders of Batebit Artesania Digital, a space for the conception and development of musical interfaces and interactive installations. He is part of the music and technology research group, MusTIC, at CIn-UFPE. He has participated in and presented works at various events at the intersection of art and technology, such as Recife: The Playable City, International Festival of Electronic Language (FILE), Digital Art Festival (FAD), Continuum, Interactivos. In 2014 and 2018, he was awarded the Itaú Cultural Rumos for the development of the musical app Tocada and the prototyping toolkit Probatio.

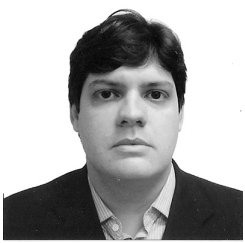

**Vanilson Burégio** holds a bachelor's, master's, and doctoral degree in Computer Science from the Federal University of Pernambuco. He began his career as a professor in 2001 and has been teaching various Software Engineering disciplines in undergraduate, postgraduate, and specialization courses at different institutions such as UNIBRATEC, CIn-UFPE, CESAR.EDU, AESO, and FBV. Currently, he is an Associate Professor at the Department of Statistics and Informatics (DEINFO) at the Federal Rural University of Pernambuco (UFRPE). In his professional experience, he worked as a software architect on information systems for the Brazilian Federal Revenue Service (at SERPRO), financial and tax administration systems for the State of Pernambuco (in collaboration with SEFAZ-PE and the Information Technology Agency of Pernambuco - ATI). Vanilson also spent six years at C.E.S.A.R. as a systems engineer on various large-scale projects, primarily using Java technology, and served as a technical leader in industrial projects with a focus on software reuse. Currently, he is involved in various national and international collaborative research groups, including the National Institute of Science and Technology for Software Engineering (INES). Vanilson is the Editor-in-Chief of "The International Journal of Information Technology and Web Engineering" and serves as a reviewer for national and international events. His research interests encompass social machine engineering, Enterprise 2.0, Internet of Things, and smart cities.

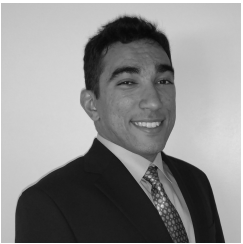
**Francisco Erivaldo** holds a Postdoctoral degree in Computer Science from the University of São Paulo (2021), a Ph.D. in Electrical Engineering from Oklahoma State University (2020), a Master's degree in Electrical Engineering from the State University of Campinas (2014), and a Bachelor's degree in Industrial Mechatronics Technology from the Federal Institute of Ceará (2012).

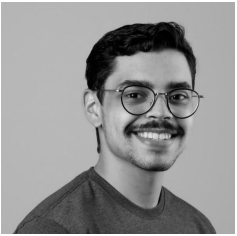
**Daniel Moraes Costa Andrade** earned his Bachelor's degree in Information Systems from the Federal University of Pernambuco (UFPE). He actively contributed to a research project focused on Generative AI and Software Development at CESAR Brazil. From 2020 to 2021, Mr. Andrade served as an Assistant Professor for the Discrete Mathematics course at UFPE. His diverse range of interests encompasses data analysis, data engineering, artificial intelligence, machine learning, big data, NLP, and information systems.

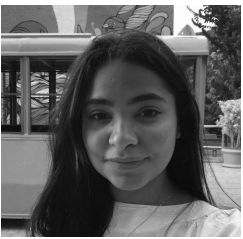
**Kailane Felix** is an undergraduate student in Computer Engineering at the Center of Informatics (CIn) at the Federal University of Pernambuco (UFPE) and currently works as a data scientist at Neurotech. Is an enthusiast of AI/ML and is dedicated to developing intelligent and creative solutions with data.

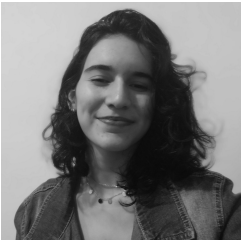

**Nathalia Barbosa** is an undergraduate student in Computer Engineering at the Center of Informatics (CIn) at the Federal University of Pernambuco (UFPE). Her main areas of interest comprise artificial intelligence, software engineering, and hardware. As part of her undergraduate studies, she conducted research, with a specific focus on developing a prototype of the RISC-V processor on FPGA. Currently, she is actively involved in research projects that explore the intersection between generative artificial intelligence and software engineering, particularly in investigating semantic merge conflicts.

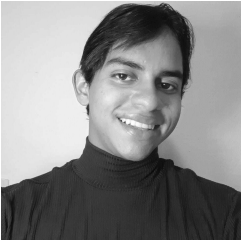

**Pedro Lucas da Silva Lucena** is an undergraduate student in Computer Engineering at the Center of Informatics (CIn) at the Federal University of Pernambuco (UFPE).

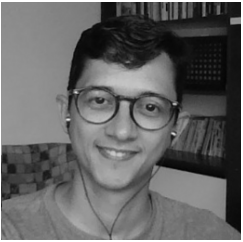

**César França** is PhD in Computer Science, and currently works as Head of Knowledge at CESAR, being responsible for engaging organizational efforts in the most promising investments, partnerships and research areas. At Cesar School, he is also faculty member in the Professional Master's and PhD programs in Software Engineering (http://cesar.school/), and teaches in some of the executive training programs. With extensive experience in Empirical and Experimental Software Engineering, he leads a Research Group on Social and Human Aspects in Software Engineeging (GENTE). He is also Professor of Software Engineering in the Department of Computing at the Federal Rural University of Pernambuco (UFRPE) (http://dc.ufrpe.br/) since 2015.